# Computation of Maximal Resolution of Copy Number Variation on a Nanofluidic Device using Digital PCR

Simant Dube*, Alain Mir, Robert C. Jones, Ramesh Ramakrishnan, Gang Sun

Fluidigm Corporation
7000 Shoreline Court, Suite 100
South San Francisco, CA 94080
*Email: simant.dube@fluidigm.com

## ABSTRACT

*Copy Number Variations (CNVs) of regions of the human genome are important in disease association studies. It was recently shown how one can perform statistical analysis of CNV in a DNA sample utilizing a nanofluidic biochip, known as the digital array. This chip utilizes integrated channels and valves that partition mixtures of sample and reagents into 765 nanovolume reaction chambers. However, when the concentration of the input target molecules is large, a chamber could have multiple molecules. In this recent work, it was shown how one can accurately estimate the true concentration of the molecules in the DNA sample and then determine the ratios of different sequences along with statistical confidence intervals on these estimations. The goals of this paper are two fold. First, we utilize this mathematical analysis by presenting optimal number of positive chambers needed to obtain tightest confidence interval. This leads to computation of maximum number of copies which can be distinguished using the digital array which gives its resolution in terms of its ability to determine CNV. Second, we demonstrate the usefulness of the mathematical analysis to solve an important real-world problem of determination of the copy number of X chromosome as our example application.*

## 1. Introduction

**Digital PCR and Digital Array**
Digital PCR conventionally utilizes sequential limiting dilutions of target DNA, followed by amplification using the polymerase chain reaction (PCR) [**1**, **2**, **3**]. As a result, it is possible to quantitate single DNA target molecules. We utilize the digital array, which is a novel nanofluidic biochip [**2, 3**,**12**] where digital PCR reactions can be performed by partitioning DNA molecules, instead of diluting them. This chip utilizes integrated channels and valves that partition mixtures of sample and reagents into 765 nanolitre volume reaction chambers, see Figure 1. The chip is then thermocycled and imaged on Fluidigm's BioMark real-time PCR system and the positive chambers that originally contained 1 or more molecules can be counted by the digital array analysis software.

**Copy Number Variation**
Copy number variations (CNVs) are the gains or losses of genomic regions which range from 500 bases on upwards in size. Whole genome studies have revealed the presence of large numbers of CNV regions in human and a broad range of genetic diversity among the general population [**4**, **5**]. CNVs have been the focus of many recent studies because of their roles in human genetic disorders [**6**].

Current whole-genome scanning technologies use array-based platforms (array-CGH and high-density SNP microarrays) to study CNVs [**7**,**8**]. They are high throughput but lack resolution and sensitivity.

CNV determination on the digital array is based upon its ability to partition DNA sequences. Given the number of molecules per panel and the dilution factor, the concentration of the target sequence in a DNA sample can be accurately calculated. In a multiplex PCR reaction with 2 or more assays, multiple genes can be quantitated simultaneously and independently.

## X Chromosone Aneuploidy

Humans typically have one only pair of sex chromosomes (Chr.) in each cell, see [**9**]. Females have two X Chr. and males one X and one Y. Abnormal X Chr. copy number - aneuploidy- is also remarkably common, ranging in incidence from: 1/2500 for 1 X Chr. in Turners syndrome; 1/750 in Klinefelters syndrome (2 extra X Chr.) and 1/1000 in Triple X syndrome (3 extra X Chr). The methods for determination of X Chr. aneuploidy include comparative genome hybridization CGH [**10**] and microarray based molecular inversion probe technology [**11**]. However, these later methods are burdened by technical complexity, and require a high level of hands-on technical involvement. Moreover, they show limited linearity and relatively low ability to distinguish between small, but biologically relevant copy number variations.

## Primary Contribution of this Paper

The copy number variation problem can be stated as follows. Given two counts $h_1$ and $h_2$ of positive chambers for two genes in a digital array panel, how can one estimate a ratio of true concentrations $r = \lambda_1 / \lambda_2$ of the two genes and a confidence interval $[r_{Low}, r_{High}]$ on the estimation?

The problem was solved recently in [**12**], see public access article at URL *http://www.plosone.org/doi/pone.0002876*, in which a mathematical framework was first derived to calculate the true concentration of molecules from the observed positive reactions in a panel. It was then shown how one can perform statistical analysis to find the 95% confidence intervals of the true concentrations and the ratio of two concentrations in a CNV experiment using the digital array with multiplex PCR.

In this paper, we first briefly show the distinction between the definitions of statistical confidence interval and Bayesian confidence interval. Then we point out that how these two intervals are mathematically identical if one does not have any prior information about the concentration of molecules. We then show how the statistical uncertainty varies with the number of positive chambers in the digital array which can act as guidelines for optimal experiment design. This allows us to discover the resolution of the digital array in terms of its ability to distinguish copy numbers which is the primary goal of this paper.

As our biological example application, we analyze separate DNA samples containing 1, 2, 3, 4 or 5 copies of the X Chr. We perform data analysis and obtain 95% confidence intervals on the number of copies.

## 2. Prior Results

DNA quantitation in the digital array is based on the partitioning of a PCR reaction into an array of several hundreds or even few thousands of chambers or wells. One panel of the digital array consists of 765 chambers and one can use up to 12 panels at a time. If the number of molecules is large, then there is greater probability of several molecules being in the same chamber, and therefore the number of positive chambers would be significantly lower compared with the number of molecules in the chambers, see **[12]**. The true concentration $\lambda$, the number of molecules per chamber, is an unknown population parameter of the DNA sample. If a chamber gets one or more molecules, that is, if it gets a hit and is therefore positive, then it constitutes success in the sense of Bernoulli trials. Let the probability of success be $p$.

The relationship between $\lambda$ and $p$ is given by

$$\lambda = -\ln(1-p)$$
$$p = 1 - e^{-\lambda}$$

For proof, see **[12]**. Also see Table 1 for list of formulas needed to obtain confidence intervals on the estimation of concentration and ratio of concentrations. We use the standard hat notation to denote sample estimators of population parameters. For example, $\hat{p}$ and $\hat{\lambda}$ denote the estimators of $p$ and $\lambda$, respectively.

The formula given in Table 1 for estimation of ratio $r = \lambda_1/\lambda_2$ of the concentration of two genes looks complicated but has an easy geometric interpretation, see **[12]**.

## 3. Results and Discussion

In this paper we continue the work in **[12]**.

### Bayesian Confidence Interval

The statistical confidence interval as derived in **[12]** assumes that the population parameter $\lambda$ is a fixed constant. An alternative approach, which gives good results depending on the question one is trying to answer, can be formulated in which $\lambda$ is a variable and has a probability distribution. The confidence interval obtained using this method is called Bayesian confidence interval or *credible interval*, see [**13**]. How is this credible interval going to be related to statistical confidence interval as derived in **[12]**?

The answer to this question, which has been known [**13**], is that if one assumes uniform distribution for $\lambda$ then the statistical confidence interval and Bayesian credible interval will be identical. For proof and a detailed discussion, see [**13**]. Such confidence interval captures the uncertainty due to possibility of different number of molecules giving same number of positive chambers, and the uncertainty caused by the fact that a small amount of DNA was pipetted out of a biological sample idealized as an infinite universe. For a related but different question of determining Bayesian interval just for one particular experiment which captures only the first uncertainty due to the possibility of different number of molecules without considering the sampling error, see [**14**]. We conjecture that a generalization of this Bayesian approach to include sampling error will lead to identical results.

### Optimal Number of Hits for Absolute Quantitation

Given uncertainty in estimation of $\lambda$, the concentration of DNA molecules, a natural question arises about the number of hits which will give least uncertainty.

In other words, the experimental biologist would be interested in knowing if there is an optimal dilution factor $d$, which changes the concentration from $\lambda$ to $d \times \lambda$, for smallest 95% confidence interval.

Suppose for $\lambda$ and $d \times \lambda$ we get confidence intervals $[a,b]$ and $[a',b']$, respectively. From estimated diluted concentration we can estimate the original concentration by dividing it by $d$ and therefore the resulting 95% confidence interval will be $\left[\dfrac{a'}{d}, \dfrac{b'}{d}\right]$. Now if

$$\frac{b'}{d} - \frac{a'}{d} < b - a$$

then dilution factor $d$ helps in reducing the uncertainty.

Since the above is equivalent to

$$\frac{b'}{d \times \lambda} - \frac{a'}{d \times \lambda} < \frac{b}{\lambda} - \frac{a}{\lambda}$$

one has to simply compute the ratio $(b-a)/\lambda$ for all values of $\lambda$ and find the minimum, see Figure 2 (where we have converted $\lambda$ into number of hits on X axis, and Y axis is in %).

### Resolution of the Digital Array

In the same way, given a ratio $r = \lambda_1/\lambda_2$ how could one dilute the samples so that one achieves smallest 95% confidence interval in the estimation of ratio of concentrations and therefore in the estimation of the CNV?

One has to simply try different dilution values $d$ for the ratio $r = (d \times \lambda_1)/(d \times \lambda_2)$ and find the value which gives smallest 95% confidence interval as given by the formula in
Table 1. In Figure 3 we show how the confidence interval varies with the total number of molecules of the reference gene. Since we assume that number of copies of reference gene is 2, one can compute the optimal confidence interval $[a,b]$ for different number of copies of test (target) gene for one panel with given number of chambers $C = 765$. Furthermore, we can now compute optimal confidence interval length any $n$ number of panels with number of chambers being $C_n = n \times 765$. It can be seen from Figure 4 that the digital array can distinguish 13 copies from 12 copies if all the 12 panels (containing 9180 chambers) are used.

Based on the numerical results, for any number of chambers $C_n$, it can be observed that the length of confidence interval approximately decreases by the square root of the number of chambers (or, the number of panels) which we conjecture can be proved analytically using Taylor series approximation as done in **[12]** to derive expectation of $\lambda$.

### X Chromosome Copy Number

See Figure 5 for results on determination of copies of X Chr. We used four different assays described in detail later in section on Materials and Methods and performed a pooled analysis by adding the hits from different assays all together in a total of $4 \times 765$ chambers. Clearly we are able to estimate the number of copies with a high degree of accuracy.

## 4. Materials and Methods
**DNA samples:**

DNA from cell lines containing either 1, 2, 3, 4 or 5 copies of the X Chr. were obtained from the Coriell Institute for Medical Research (Camden, NJ). Four separate X Chr. test TaqMan® reactions (FAM123B, G6PD, SMS and YY2; FAM labeled probe) were amplified in the presence of a single copy targeting, VIC labeled "reference" sequence. The reference sequence lies between genes encoding Human H1 (RNAse P) and a polyA-polymerase on human Chr. 14.

**Method:**
60 ng of X Chr. copy number variant DNA was amplified in a PreAmp Master Mix (Applied Biosystems), containing all primers and probes, cycled 10 times (95 ºC/15 secs., 60ºC/120 secs.) and diluted 1/15,000 in water. 1.5 µL was mixed with PCR master mix, DA loading buffer (Fluidigm), separate test and reference specific primer-probe mixes. 8-from-10 µL was loaded into separate panels of a Fluidigm Digital PCR array. 40 cycles of PCR amplification was performed (95ºC/15 secs. 60ºC/60 secs.) and fluorescence intensity collected.

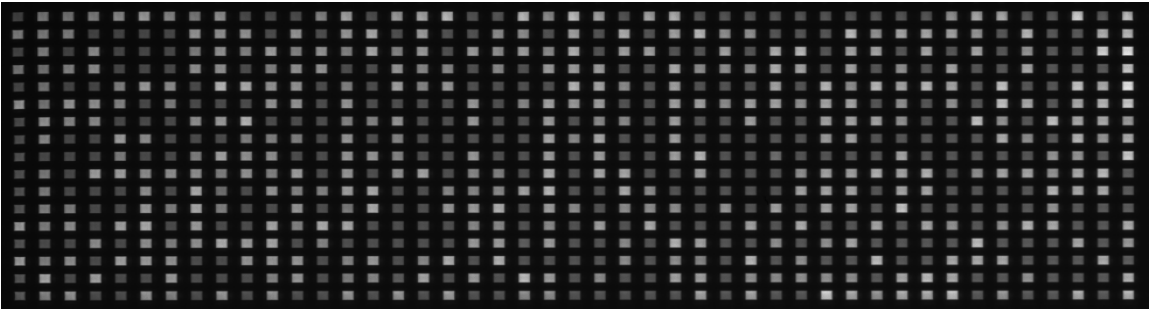

**Figure 1**: One panel of the Digital Array. Each chamber contains 6 nanolitre of DNA sample and reagent mixture. Each panel has 765 chambers and in total there are 12 panels. If there is one or more molecules in a chamber then it is marked as a "hit" (or, positive chamber). From the count of hits, we estimate the concentration of molecules and statistical confidence interval using formulas shown in Table **1**.

$$\hat{p}_1 = \frac{H_1}{C},\ \hat{p}_2 = \frac{H_2}{C},\ S_1 = \sqrt{\frac{\hat{p}_1(1-\hat{p}_1)}{C}},\ S_2 = \sqrt{\frac{\hat{p}_2(1-\hat{p}_2)}{C}}$$

$$\hat{p}_{1,Low} = \hat{p}_1 - 1.96 S_1,\ \hat{p}_{1,High} = \hat{p}_1 + 1.96 S_1,\ \hat{p}_{2,Low} = \hat{p}_2 - 1.96 S_2,\ \hat{p}_{2,High} = \hat{p}_2 + 1.96 S_2$$

$$\hat{\lambda}_1 = -\ln(1-\hat{p}_1),\ \hat{\lambda}_{1,Low} = -\ln(1-\hat{p}_{1,Low}),\ \hat{\lambda}_{1,High} = -\ln(1-\hat{p}_{1,High})$$

$$\hat{\lambda}_2 = -\ln(1-\hat{p}_2),\ \hat{\lambda}_{2,Low} = -\ln(1-\hat{p}_{2,Low}),\ \hat{\lambda}_{2,High} = -\ln(1-\hat{p}_{2,High})$$

$$H_T = \hat{\lambda}_{1,High} - \hat{\lambda}_1,\ H_B = \hat{\lambda}_1 - \hat{\lambda}_{1,Low},\ W_R = \hat{\lambda}_{2,High} - \hat{\lambda}_2,\ W_L = \hat{\lambda}_2 - \hat{\lambda}_{2,Low}$$

$$\hat{r} = \frac{\hat{\lambda}_1}{\hat{\lambda}_2},\ \hat{r}_L = \frac{\hat{\lambda}_1 \hat{\lambda}_2 - \sqrt{\hat{\lambda}_1^2 \hat{\lambda}_2^2 - (H_B^2 - \hat{\lambda}_1^2)(W_R^2 - \hat{\lambda}_2^2)}}{\hat{\lambda}_2^2 - W_R^2},\ \hat{r}_H = \frac{\hat{\lambda}_1 \hat{\lambda}_2 + \sqrt{\hat{\lambda}_1^2 \hat{\lambda}_2^2 - (H_T^2 - \hat{\lambda}_1^2)(W_L^2 - \hat{\lambda}_2^2)}}{\hat{\lambda}_2^2 - W_L^2}$$

**Table 1:** Given number of chambers $C$ and counts $H_1$ and $H_2$ of the positive chambers in a digital array for the target gene and the reference gene, respectively, list of formulas needed to analyze copy number variation. See [12] for derivations of these formulas.

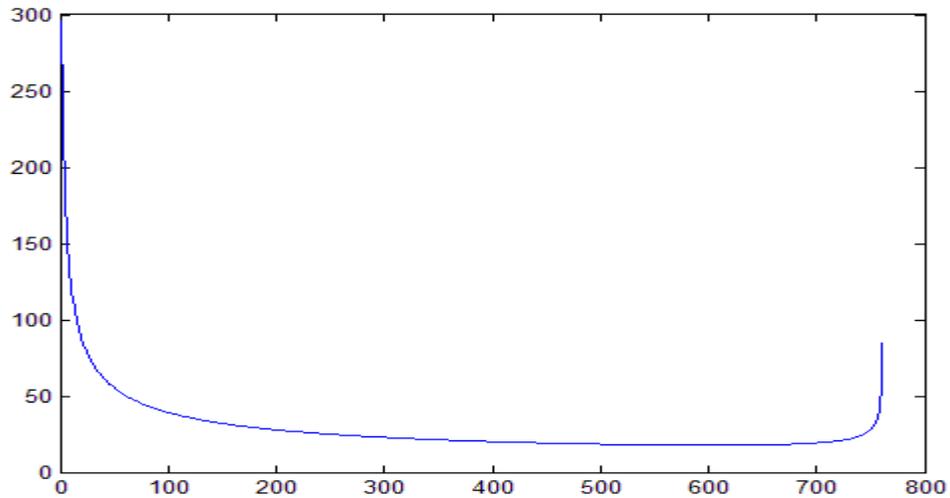

**Figure 2:** Graph showing ratio of the length of the confidence interval to concentration in % on Y axis versus the number of positive chambers (hits) on X axis. One can infer that a wide range of number of hits from 200 to 700 gives smallest possible uncertainty in the estimation of the concentration of the gene.

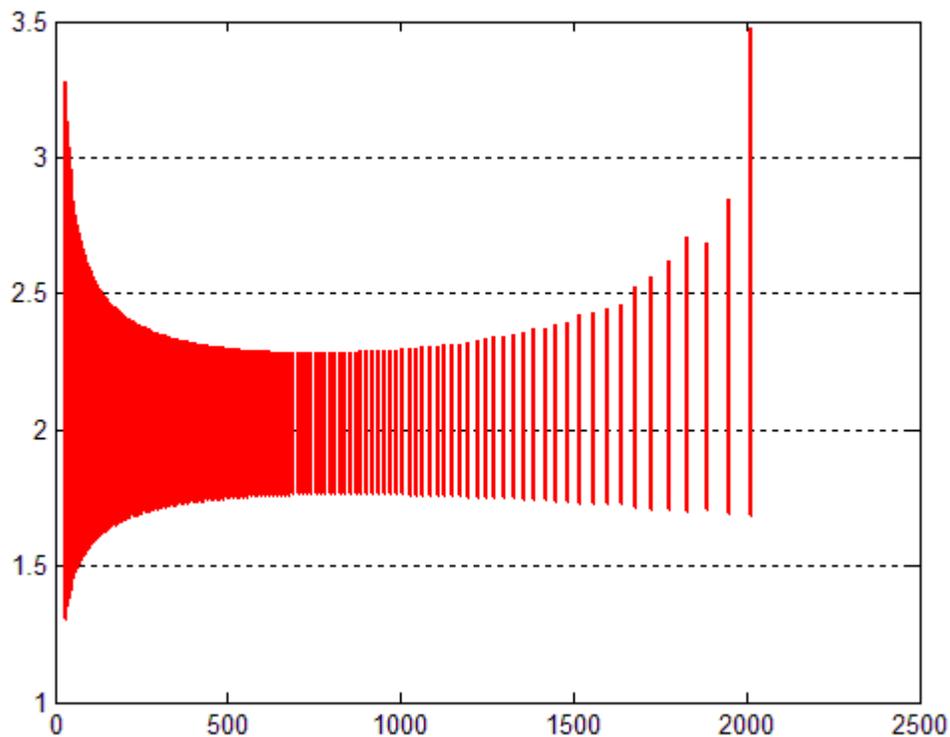

**Figure 3:** Graph showing how for a CNV ratio of 2 (i.e. test gene has 4 copies and reference gene has 2 copies) and one panel of 765 chambers, the uncertainty in the estimation of the ratio (Y axis) varies with the total number of molecules of the reference gene in the panel (X axis). One can infer that there is a wide range of dilution factors which give close to smallest possible uncertainty in the estimation of the CNV. The uncertainty is defined as the 95% confidence interval.

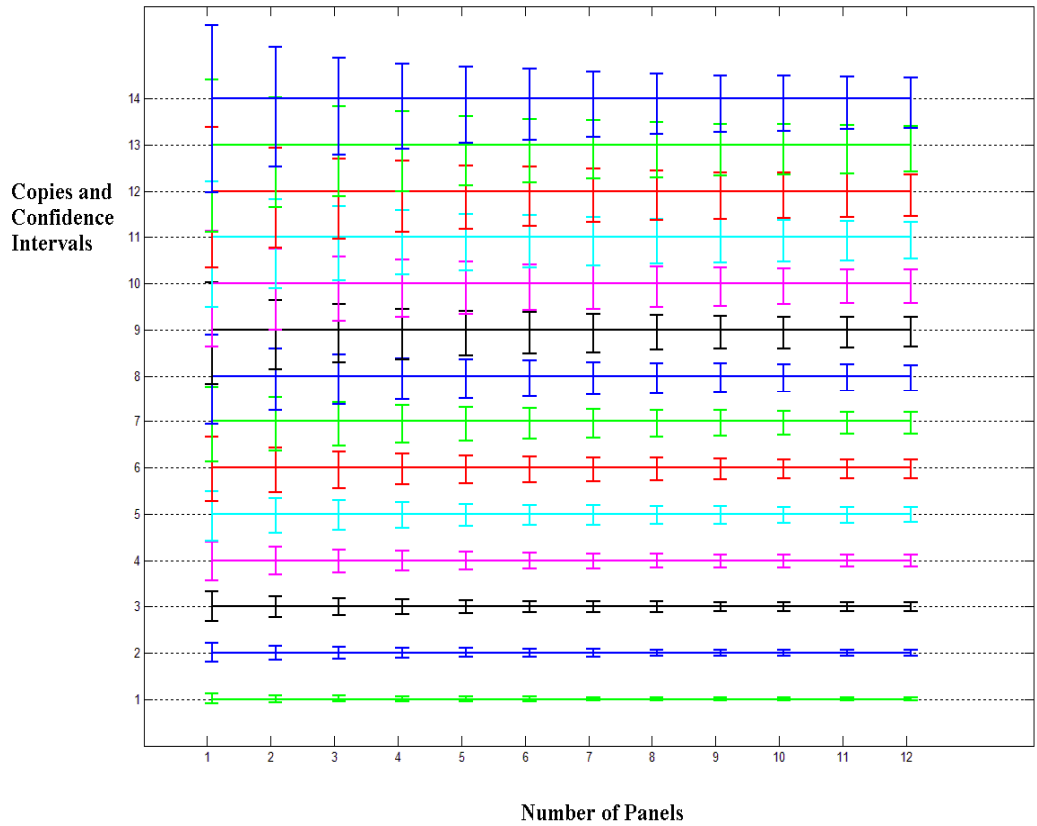

**Figure 4**: Graph showing resolution of the digital array in terms of its ability to distinguish copies of genes. X axis shows the number of panels used. (Each panel has 765 chambers.) Y axis shows the copies of the target gene and smallest possible 95% confidence interval.

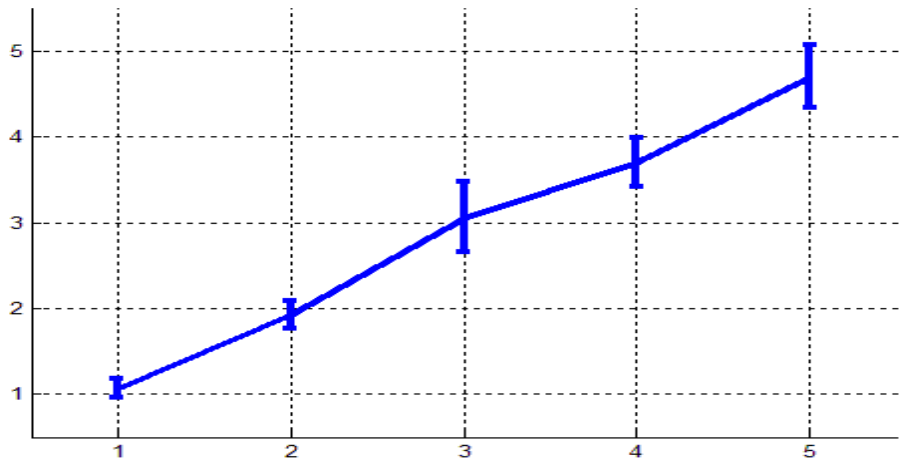

**Figure 5:** Graph showing estimated number of copies of X chromosome along with 95% confidence intervals. X axis gives the known number of X Chr. copies. Y axis gives the estimated number of X Chr. copies and the 95% confidence intervals. In all cases we are able to accurately estimate the number of copies.